\documentstyle[aps,prb,epsf,multicol]{revtex}

\begin{document}
\setlength{\topmargin}{-1.5cm}

\title{Density Matrix Renormalization Group of Gapless Systems}

\author{Martin Andersson, Magnus Boman and Stellan \"Ostlund}

\address{Institute of Theoretical Physics, \\ Chalmers University of 
Technology and G\"{o}teborg University, \\ S-412 96 G\"{o}teborg, Sweden}

\date{\today}

\maketitle

\begin{abstract}
We investigate convergence of the density matrix renormalization group 
(DMRG) in the thermodynamic limit for gapless systems. Although the DMRG 
correlations always decay exponentially in the thermodynamic limit, the 
correlation length at the DMRG fixed-point scales as $\xi \sim m^{1.3}$, 
where $m$ is the number of kept states, indicating the existence of 
algebraic order for the exact system. The single-particle excitation 
spectrum is calculated, using a Bloch-wave ansatz, and we prove that the 
Bloch-wave ansatz leads to the symmetry $E(k)=E(\pi -k)$ for 
translationally invariant half-integer spin-systems with local interactions.
Finally, we provide a method to compute overlaps between 
ground states obtained from different DMRG calculations.
\end{abstract}
\bigskip

%%%%%%%%%%%%%%%%%%%%%%%%%%%%%%%%%%%%%%%%%%%%%%%%%%%%%%%%%%%%%%%%%%%%%%
%%%%%%%%%%%%%%%%%%%%%%%%%%%%%%%%%%%%%%%%%%%%%%%%%%%%%%%%%%%%%%%%%%%%%%

\begin{multicols}{2}
\narrowtext

\section{Introduction}\label{intro}

Since White constructed the density matrix renormalization group (DMRG) 
technique about five years ago\cite{white1}$^,$\cite{white2}, numerical 
renormalization techniques have become very useful. The DMRG has by now 
been applied to a wide range of different problems beyond quantum spin 
systems, where it was originally used. Today, people use it to compute, 
for instance, properties of two-dimensional classical lattice 
systems\cite{nishino}, thermodynamics of one-dimensional quantum 
systems\cite{bursill} etc. The DMRG and matrix product states
\cite{accardi}$^-$\cite{klumper2} have proven to 
be computationally very efficient and to determine properties of many 
systems with unusually high accuracy. 

This paper aims at a better understanding of the underlying structure 
and the fundamental limitations of the DMRG. It has been reported that the
DMRG is less accurate for gapless systems than for gapped systems
\cite{drzewinski}$^,$\cite{legeza}.
This has motivated us to analyse the DMRG of gapless systems in more detail,
which we do in two steps: first, we investigate the correlation functions
(which pertain to the wavefunction), and second, we study the excitation
spectrum. We have choosen to study free fermions on a
one-dimensional lattice as a paradigm of gapless systems.

In this work, we show that the DMRG of a gapless system converges, for 
each choice of the number of kept states, to a
fixed-point. The corresponding correlation functions are calculated by using
an eigenvalue technique\cite{ostlund}$^,$\cite{rommer}, and the results show 
that this fixed-point describes with increasing accuracy the exact system.
In particular, we address the question of how the DMRG
handles algebraic correlations (infinite correlation lengths), which
is characteristic of gapless systems, in contrast to gapped systems having 
finite corrrelation lengths and determine a scaling formula
for how the DMRG correlation length depends on the number of kept
states. In DMRG applications, this scaling formula may serve as a
guide for how many states that is neccessary to keep in order to
accuratly calculate correlations. 
We perform calculations for the particle-hole and the
density-density correlation functions. In addition, we derive conditions
for which types of operators that can give truly long ranged DMRG 
correlations. 
By introducing a gap, we also investigate how the DMRG correlation lengths
change as the critical point (free fermions) is approached.

A matrix product Bloch-wave ansatz has been proposed for describing the excited
states \cite{ostlund}$^,$\cite{rommer} of a system. In this paper, we have 
used the Bloch-wave ansatz to calculate the excitation spectrum. In 
particular, we look at the spectrum close to the Fermi points, where the gap 
closes. Furthermore, it was recently shown\cite{kladko} that many 
half-integer spin systems have the symmetry $E(k)=E(\pi -k)$. We prove that 
this symmetry is inherent also in the Bloch-wave ansatz.

In DMRG calculations, it is important to check convergence with
respect to the only source of error (except from round off errors), namely 
the truncation of the Hilbert space. A commonly used measure of the 
truncation error is the truncation of the density matrix\cite{white1}$^,
$\cite{white2}. However, that measure is algorithm dependent and therefore 
not universal. It would be desirable to instead calculate and use the 
overlap between states from different DMRG calculations as a measure. 
The problem is then that the states will refer to differently renormalized 
Hilbert spaces. In this work, we demonstrate how the matrix product formalism 
can be used to handle this problem.

The organization of the paper is as follows: the system we have
studied is defined in Section \ref{fermions}; a brief introduction
to matrix product states and the Bloch-wave ansatz is given in Section
\ref{matprod}; we decribe how we calculate correlation functions
and overlaps in Section \ref{Met: corr} and \ref{Met: overlap}; computational
methods are outlined in Section \ref{Met: Comp}.
The results are presented and discussed in Section \ref{Res:}:
convergence of the projection operator (to a fixed-point) and of the 
ground state is demonstrated in Section \ref{Res: conv};
correlation functions and the scaling formula are treated
in Section \ref{Res: corr}; Appendix \ref{fhatspec} contains an important 
result for the determination of correlation lengths in a fermionic system; 
conditions for true long range correlations are derived in Appendix 
\ref{longrange}; the excitation spectra is presented in Section 
\ref{bloch} and a proof of the symmetry of the spectra is given in Appendix 
\ref{blochproof}. Finally, the results and conclusions are summerized in 
Section \ref{conc}.

%%%%%%%%%%%%%%%%%%%%%%%%%%%%%%%%%%%%%%%%%%%%%%%%%%%%%%%%%%%%%%%%%%%%%%
%%%%%%%%%%%%%%%%%%%%%%%%%%%%%%%%%%%%%%%%%%%%%%%%%%%%%%%%%%%%%%%%%%%%%%

\section{Methodology}\label{methods}

\subsection{Hamiltonian}\label{fermions}

We have studied a system of non-interacting spinless fermions on a
one-dimensional lattice. The Hamiltonian is:
\begin{equation}
H=-\frac{t}{2}\sum_{j=1}^{N}[c^{\dagger}_{j}c_{j+1}+\hbox{\rm h.c.}]+
\epsilon\sum_{j=1}^{N}(-1)^{j}c_j^{\dagger}c_j,
\end{equation}
where $N$ is the size of the lattice, $c_j^{\dagger}$ creates a fermion on
site $j$, and $t$ is the hopping amplitude\cite{hamsign} ($t=2$
throughout this work). We have added a staggered on-site potential
$\epsilon$ to the lattice
since we want to compare DMRG of the gapless system ($\epsilon=0$) to that
of gapped systems ($\epsilon\neq 0$).
The simple Hamiltonian gives us the advantage of having access to exact 
solutions when evaluating the DMRG. In the remaining part of this section, 
we will state exact results used in our analysis.
 
The Hamiltonian is particle-number conserving and is invariant under the
transformation $c_j\rightarrow (-1)^jc_{j+1}^{\dagger}$ for all $\epsilon$.
 In addition, when 
$\epsilon=0$, the Hamiltonian has particle-hole symmetry, i.e. it is 
invariant under the transformation
$$
c_j\rightarrow (-1)^jc_j^{\dagger}.
$$

We will only consider chains of length $N=4n+2$ in order to have 
a unique ground state, which corresponds to a half filled
system\cite{energydensity}. The gap between the valence band and the
conduction band at the Fermi points is $2|\epsilon|$.
The correlation functions $C(l)$ 
decay algebraically for the gapless system and exponentially for a gapped
system.
When $\epsilon=0$ we have
\begin{equation}
C_{ph}(l)=\langle c_j^{\dagger}c_{j+l}\rangle = \frac{1}{\pi l}\sin\pi l/2,
\label{eq:partholecorr}
\end{equation}
and
\begin{equation}
C_{dd}(l)=\langle n_jn_{j+l}\rangle -\langle n_j\rangle\langle n_{j+l}\rangle
= -\frac{1}{\pi^2l^2}\sin^2\pi l/2 
\label{eq:densitycorr}
\end{equation}
for the particle-hole and density-density correlation functions 
respectively, where $n_j=c_j^{\dagger}c_j$.

When $\epsilon\neq 0$, the correlation length for the ground state
particle-hole correlation function can be calculated
analytically\cite{psi anal} to be
\begin{equation}
\xi_{ph} (t,\epsilon)=\frac{1}{\ln \Bigl[ \epsilon /t+
\sqrt{1+(\epsilon /t)^2}\Bigr]}.
\label{eq:phcorrlength}
\end{equation}
Similarly, the exact density-density correlation length is given by 
$\xi_{dd}=\xi_{ph}/2$.

Finally, we note that there is a well known
connection to spin-systems. We associate a spin-1/2 with each site in the 
lattice and consider an occupied site as spin up and an empty site as spin 
down. Using the spin raising and lowering operators we may rewrite the 
Hamiltonian (after a Jordan-Wigner transformation) as
\begin{eqnarray}
H &=& -\frac{t}{2}\sum_{j=1}^N[S_j^+S_{j+1}^-+\hbox{\rm h.c.} ]
+\epsilon\sum_{j=1}^N
(-1)^j(S_j^z+1/2) \cr
  &=& -t\sum_{j=1}^N[S_j^xS_{j+1}^x+S_j^yS_{j+1}^y]+\epsilon\sum_{j=1}^N
(-1)^jS_j^z,
\label{eq:spinham}
\end{eqnarray}
where as usual $S^\pm = S^x\pm iS^y$. The number operator $n_j$ in 
fermionic terminology is identified as $S_j^z+1/2$ in the spin terminology.

%%%%%%%%%%%%%%%%%%%%%%%%%%%%%%%%%%%%%%%%%%%%%%%%%%%%%%%%%%%%%%%%%%%%%%

\subsection{Matrix Product States}\label{matprod}

We refer to previous work\cite{ostlund}$^,$\cite{rommer} for a derivation 
of the ansatz and for details concerning the calculations in this section. 
A general matrix product state takes the form
\begin{equation}
|Q)_N=\sum_{\{ s_j\}}\hbox{\rm tr} \bigl[ QA[s_N]\cdots A[s_1]\bigr] 
|s_N\cdots s_1\rangle ,
\label{eq:mpansatz}
\end{equation}
where $Q$ is an $m\times m$ matrix containing the boundary conditions on 
the chain, $A[s]$ is an $m\times m$ projection matrix obtained either from 
a DMRG calculation or variationally, and $s_j$ is the quantum number 
associated with site $j$. The projection matrix $A$ contains the information 
about which states to keep when the lattice is augmented with one site. The 
number of degrees of freedom in $A$ is reduced by preservation of orthonormal 
bases: $\sum_sA[s]A^{\dagger}[s]=\openone_m$. Further reduction of the 
number of free parameters is possible by exploiting symmetries of the 
system. In our variational calculations we have used particle-hole symmetry 
and conservation of the number of particles. Generally, the projection 
matrix $A[s]$ is built up from states that form irreducible representations 
of the symmetry group of the Hamiltonian. 

In terms of the spin Hamiltonian Eq. (\ref{eq:spinham}), each factor $A[s]$ in 
Eq. (\ref{eq:mpansatz}) adds a spin-1/2, hence taking a 
half-integer ({\tt hi}) total spin into an integer ({\tt i}) total spin 
and vice versa. This implies that we can define our projection matrix $A$ 
with an off-diagonal block structure
\begin{equation}
A[s]=
\left(
\begin{array}{cc}
0	&	A_{{\tt hi}\rightarrow {\tt i}}[s]\\
A_{{\tt i}\rightarrow {\tt hi}}[s]	& 0\\
\end{array}
\right) .
\label{eq:ablock}
\end{equation}

It is convenient to introduce the following mapping, denoted 
$\widehat{\hspace{5pt}}$, from a local $s\times s$ matrix $M$ to an 
$m^2\times m^2$ matrix $\widehat{M}$:
\begin{equation}
\widehat{M}=\sum_{s',s}M_{s',s}A^*[s']\otimes A[s].
\end{equation}
Just using the block structure of $A[s]$, one can show\cite{evalpair} that 
the eigenvalues of an operator $\widehat{M}$ appear in pairs $\pm \lambda$. 
We will frequently interpret the eigenvectors of $\widehat{M}$ (of length 
$m^2$) as matrices of size $m\times m$.

The matrix $\widehat{1}$, i.e. the $\widehat{\hspace{5pt}}$-image of the 
identity matrix, plays an important role in the theory. Since $\widehat{1}$ 
is guaranteed to have an eigenvalue $1$, due to the orthonormalization 
condition, there also exists an eigenvalue $-1$. The block structure of 
the projection matrix also implies that there may occur eigenvalues due 
to mixing of the integer and half-integer representations. However, these 
eigenvalues are spurious
(unphysical), in the sense that they do not affect the correlation 
functions, and can be removed simply by working with two different $A$ 
matrices formed of the two off-diagonal blocks in Eq. (\ref{eq:ablock}). For 
this reason we leave these spurious eigenvalues aside in the subsequent 
discussion.

For a translationally invariant system Eq. (\ref{eq:mpansatz}) can be 
generalized to a Bloch-wave ansatz:
\begin{eqnarray}
|Q,k)_N&=&\sum_{j,\{ s\}}e^{ijk}\hbox{\rm tr} \bigl[ A[s_N]\cdots A[s_{j+1}]Q
\cdots A[s_1]\bigr]\cr
&&\times |s_N\cdots s_1\rangle ,
\label{eq:bloch}
\end{eqnarray}
where $k$ is the momentum.

The ground state of our model is translationally invariant. This implies 
that the matrix $Q$ in Eq. (\ref{eq:mpansatz}) should satisfy $[Q,A[s]]=0$ 
for all $s$, or equivalently, $Q$ must be a generalized right 
eigenvector\cite{nonsymmetry} of $\widehat{1}$ with corresponding eigenvalue 
$1$ (provided that this eigenvalue is non-degenerate). This means that 
$Q\sim\openone_m$ and hence our ground state ansatz takes the form,
\begin{equation}
|1)_N=\sum_{\{ s_j\}}\hbox{\rm tr} \bigl[ A[s_N]\cdots A[s_1]\bigr] 
|s_N\cdots s_1\rangle .
\label{eq:gndstate}
\end{equation}

Note that if we use a different sign convention in the 
Hamiltonian\cite{hamsign}, the ground state would have momentum $\pi$ 
and hence we would have to choose a $Q$ that anticommutes with $A[s]$, 
$\{ Q,A[s]\} =0$ for all $s$. From this it follows that we must choose $Q$ 
as the generalized (right) eigenvector corresponding to the eigenvalue $-1$ 
of $\widehat{1}$. Let us denote this generalized eigenvector by $R$ for 
future purposes. This indicates that $R$ is associated with momentum $\pi$. 
Further evidence for this is given in connection to the discussion of the 
single particle excitation spectrum, see Section \ref{bloch}.

%%%%%%%%%%%%%%%%%%%%%%%%%%%%%%%%%%%%%%%%%%%%%%%%%%%%%%%%%%%%%%%%%%%%%%

\subsection{Correlation functions}
\label{Met: corr}

Suppose we want to compute the ground state correlation function $C(l)$ 
between two local operators $M^1_j$ and $M^2_{j+l}$ acting on sites $j$ 
and $j+l$ respectively. Since we are working with a fermionic model it 
is necessary to distinguish between local operators depending on whether 
they commute or anticommute on different sites. We refer to these operators 
as bosonic and fermionic respectively. For example, the density-density 
correlation is expressed in terms of two bosonic operators, while the 
particle-hole correlation is expressed in terms of two fermionic operators. 
We will use the superscripts $B$ and $F$ to denote bosonic and fermionic 
operators respectively. For bosonic operators the correlation function is 
given by
\begin{equation}
C(l)=\langle M^{B_1}_jM^{B_2}_{j+l}\rangle = (1|1)^{-1}\hbox{\rm tr} \bigl[ 
\widehat{M^{B_1}}
\widehat{1}^{l-1}\widehat{M^{B_2}}\widehat{1}^{N-l-1}\bigr] .
\label{eq:expecboson}
\end{equation}
If we instead are interested in a correlation function between fermionic 
operators, we have to keep track of the number of fermions between the 
sites $j$ and $j+l$ in order to get the phases correct. Defining $F$ as a 
diagonal matrix with diagonal elements $(-1,1)$, we find the expression to 
be
\begin{equation}
C(l)=\langle M^{F_1}_jM^{F_2}_{j+l}\rangle = (1|1)^{-1}\hbox{\rm tr} 
\bigl[ \widehat{M^{F_1}}
\widehat{F}^{l-1}\widehat{M^{F_2}}\widehat{1}^{N-l-1}\bigr] .
\label{eq:expecfermion}
\end{equation}

Eqs. (\ref{eq:expecboson}) and (\ref{eq:expecfermion}) imply\cite{rommer} 
that in general a correlation function takes the analytical form
\begin{equation}
C(l)=\sum_{i=1}^{m^2}\alpha_i\hbox{\rm sgn}\lambda_i^l\exp [-l/\xi_i],
\label{eq:corr}
\end{equation}
where $\xi_i=-1/\ln |\lambda_i |$, and $\lambda_i$ are the eigenvalues 
of $\widehat{1}$ or $\widehat{F}$, depending on the statistics of the 
operators. The $\alpha_i$'s are coefficients that depend on the operators 
in the correlation function. Thus, the eigenvalues of $\lambda_i$ determine 
the possible correlation lengths in the system and it is therefore important 
to investigate the spectrum of $\widehat{1}$ and $\widehat{F}$. Eigenvalues 
that fulfill $|\lambda_i|=1$ can potentially give rise to true long-range 
order. Due to normalization, $\widehat{1}$ is guaranteed to have eigenvalues 
$\pm 1$, which potentially could give long-range order in the bosonic 
correlation functions. In Appendix \ref{fhatspec} we show that the spectrum 
of $\widehat{F}$ differs from that of $\widehat{1}$ only by a factor $i$. 
Hence, fermionic and bosonic operatorss have the same set of possible 
correlation lengths, which means that $\widehat{F}$ has eigenvalues $\pm i$ 
and these can give rise to true long-range order in the fermionic correlation 
functions. Finally, we note that negative and imaginary eigenvalues 
correspond to oscillating correlation functions.

%%%%%%%%%%%%%%%%%%%%%%%%%%%%%%%%%%%%%%%%%%%%%%%%%%%%%%%%%%%%%%%%%%%%%%
 
\subsection{Overlap of DMRG states}
\label{Met: overlap}

Suppose we perform two different DMRG calculations, keeping $m_1$ and $m_2$
states respectively. The overlap $_N(1_{m_1}|1_{m_2})_N$,
where $|1_{m_2})_N$ is the normalized ground state obtained by keeping $m_2$ 
states etc., can be computed as follows,
\begin{equation}
_N(1_{m_1}|1_{m_2})_N=\hbox{\rm tr} \bigl[ \widehat{1}_{m_1,m_2}^N \bigr] ,
\label{eq:overlap}
\end{equation}
where we have defined the mixed $(m_1m_2)\times (m_1m_2)$ matrix 
$\widehat{1}_{m_1,m_2}$ as
\begin{equation}
\widehat{1}_{m_1,m_2}=\sum_sA_{m_1}^*[s]\otimes A_{m_2}[s].
\end{equation}
Note that this overlap would be difficult to compute without the matrix
product formalism since we have no mapping between the different basis states
of the two DMRG calculations due to renormalization. In contrast, all matrix 
product states are formulated in terms of the fixed 
$\{ |s_N\cdots s_1\rangle\}$ basis (rather than renormalized basis sets), 
with the projection matrix just providing the amplitudes.

Using Eq. (\ref{eq:overlap}) we find that the overlap decays exponentially as 
$\lambda^N$, where $\lambda$ is the (in absolute value) leading eigenvalue 
of $\widehat{1}_{m_1,m_2}$.

The overlap between DMRG states for different number of kept states 
gives a measure of the gain in accuracy obtained by increasing the number
of kept states. We consider this measure to be more universal and relevant 
than the usual measure used, namely the truncation of the density matrix, 
$1-\hbox{\rm tr}\rho_e$, where $\rho_e$ is the truncated density matrix. The old 
measure is algorithm-dependent and can, for instance, be made equal to 
zero\cite{white2}, despite the fact that an exact calculation is not p
erformed, simply by  using a superblock configuration $
\hbox{\framebox[1cm][c]{\scriptsize B} $\bullet$ 
\framebox[1cm][c]{\scriptsize B$^r$}}$.

%%%%%%%%%%%%%%%%%%%%%%%%%%%%%%%%%%%%%%%%%%%%%%%%%%%%%%%%%%%%%%%%%%%%%%

\subsection{Computational methods}
\label{Met: Comp}

Due to the large dimensions of $\widehat{1}_{m_1,m_2}$, namely 
$(m_1m_2)\times (m_1m_2)$, it becomes necessary to use iterative eigenvalue 
routines that require no explicit construction or storage of 
$\widehat{1}_{m_1,m_2}$. Moreover, $\widehat{1}_{m_1,m_2}$ is non-symmetric. 
We have used the Arnoldi algorithm\cite{arnoldi} to handle these problems.

Furthermore, the computations become much more efficient if we rewrite the 
operation of $\widehat{1}_{m_1,m_2}$ on an $(m_1m_2)$ vector $v$ as a matrix 
product with $v$ interpreted as an $m_1\times m_2$ matrix:
$$
\widehat{1}_{m_1,m_2}v=\sum_{s}A_{m_1}^*[s]vA_{m_2}^T[s].
$$
In this way we only need to operate with $m_1\times m_2$ matrices, and 
the eigenvalues of $\widehat{1}_{m_1,m_2}$ can easily be obtained.

\begin{figure}[tbh]
\centerline{\epsfxsize=0.6\columnwidth\epsffile{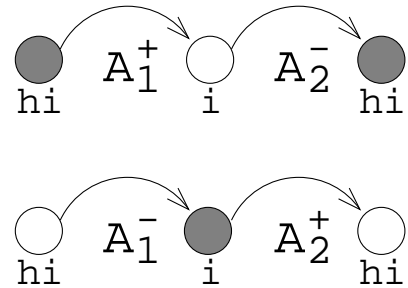}}
\bigskip
\caption[projection]{\label{fig:projection} The figure shows the four 
projection matrices needed to describe 
the system. Filled discs denote sites with positive on-site potential and 
circles denote sites with negative on-site potential, {\tt hi} denotes 
half-integer representations and {\tt i} denotes integer representations. 
Andersson et. al.}
\end{figure}

In order to calculate the projection operators $A[s]$, we have performed 
standard DMRG calculations by using a superblock of the form: 
$$
\hbox{\framebox[1.5cm][c]{\scriptsize B} $\bullet$ $\bullet$ 
\framebox[1.5cm][c]{\scriptsize B$^r$},} 
$$
the infinite lattice algorithm, and by adding a single site per iteration
to each block. When we have an on-site potential $\epsilon$ present in the 
problem we have to keep four projection matrices in order to completely 
describe the system, see Fig. \ref{fig:projection}. From these four 
matrices we form two projection operators $A^+$ and $A^-$ taking us from 
positive to negative on-site potential and vice versa. Explicitly:
$$
A^\pm[s]=\left(
\begin{array}{cc}
0	&	A^\pm_1[s]\\
A^\pm_2[s]	& 0\\
\end{array}
\right) .
$$

In each DMRG iteration we update either the ${\tt hi}\rightarrow {\tt i}$ 
or the ${\tt i}\rightarrow {\tt hi}$ matrices.

%%%%%%%%%%%%%%%%%%%%%%%%%%%%%%%%%%%%%%%%%%%%%%%%%%%%%%%%%%%%%%%%%%%%%%
%%%%%%%%%%%%%%%%%%%%%%%%%%%%%%%%%%%%%%%%%%%%%%%%%%%%%%%%%%%%%%%%%%%%%%

\section{Results and Discussion}
\label{Res:}

\subsection{Convergence of the DMRG}
\label{Res: conv}

In this section we will discuss the convergence of the DMRG.
First, we demonstrate that the DMRG projection operator of the
critical system converges to a fixed-point, and thus, justifying the 
matrix product ansatz when studying the 
thermodynamic limit of the DMRG. Second, we check the convergence of the
ground state with respect to the number of kept states by using
the overlap measure.

\begin{figure}[tbh]
\centerline{\epsfxsize=\columnwidth\epsffile{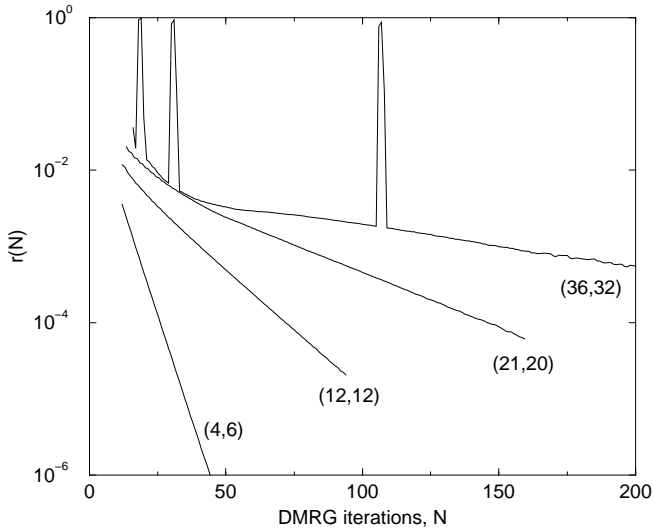}}
\caption[projection]{\label{fig:projconv}The norm 
$r(N)$ is shown 
as a function of the number of DMRG iterations, $N$, for the gapless case. 
It is clear from the figure that the DMRG projection operator converges 
with respect to $N$. Furthermore, the convergence seems to be exponential 
with respect to $N$. The peaks in the $(36,32)$ curve indicate that either 
the DMRG has changed the states kept in the Hilbert space basis or our 
sign-fixing procedure of the states has failed. The notation $(m_1,m_2)$ 
means $m_1$ states in the integer representation and $m_2$ states in the 
half-integer representation. }
\end{figure}

The fundamental assumption of the matrix product approach is that the 
projection matrix converges to a fixed-point with respect to $N$, i.e. 
$\lim_{N\rightarrow\infty}A_N[s]=A[s]$. In order to show this, we define 
the matrix norm $\|\cdot\|_{max}$ via,
\begin{equation}
\| A\|_{max}=\max_{i,j,s}|A_{i,j}[s]|.
\label{eq:matnorm}
\end{equation}
In addition, we make a consistent enumeration (with respect to quantum numbers)
and use a fixed sign convention of the states in the system blocks.
It is then easy to study the convergence of the projection operator by 
measuring the quantity $r(N)=(\| A_{N+1}-A_N\|_{max}+\| A_N-A_{N-1}
\|_{max})/2$, where $N$ is the number of DMRG iterations. 
In Fig. \ref{fig:projconv} we have shown results from such calculations.
From the figure we see that the 
convergence of the projection operator seems to be exponential with respect 
to the number of DMRG iterations and that the convergence rate decreases 
when the number of kept states is increased. We have also found that the 
ground state energy per site converges much faster than the projection 
matrices and is therefore not a good indicator on whether or not a 
fixed-point has been reached. 

\begin{tabular}{ccccc}
\hline\hline
$(m_1,m_2)$  & (4,6) & (12,12) & (21,20) & (36,32) \\ \hline
(4,6) & 0 & 0.00231 & 0.00328 & 0.00381 \\ 
(12,12) & 0.00231 & 0 & 0.000277 & 0.000644 \\ 
(21,20) & 0.00328 & 0.000277 & 0 & 0.000134 \\ 
(36,32) & 0.00381 & 0.000644 & 0.000134 & 0 \\ \hline\hline
\end{tabular}
\vspace{2mm}

{\small TABLE I. Leading eigenvalues $1-\lambda$, govering the overlap between 
different DMRG ground state wavefunctions. The notation $(m_1,m_2)$ means 
$m_1$ states in the integer representation and $m_2$ states in the 
half-integer representation.}\bigskip

In Table. I we have presented the eigenvalues that govern 
the overlap in the thermodynamic limit between ground states obtained by 
keeping different numbers of states. We have chosen to write out 
$1-\lambda$ instead of $\lambda$ since this gives a more direct measure 
of the error.\bigskip

%%%%%%%%%%%%%%%%%%%%%%%%%%%%%%%%%%%%%%%%%%%%%%%%%%%%%%%%%%%%%%%%%%%%%%

\subsection{The spectrum of $\widehat{1}$ and correlation lengths}
\label{Res: corr}

Since the DMRG projection operator converges to a fixed-point, the 
correlation functions of the DMRG in the thermodynamic limit are given 
by Eq. (\ref{eq:corr}) and the correlation lengths are determined by the 
eigenvalues of $\widehat{1}$ (the correlation lengths obtained from 
$\widehat{F}$ are identical). An analysis of the spectrum of 
$\widehat{1}$ is therefore pivotal.

We have found that the eigenvalues $\pm 1$ are non-degenerate, and that all
the other eigenvalues fulfill $|\lambda |< 1$. Only the eigenvalues $\pm 1$ 
can give rise to infinite correlations lengths. However, it turns out that 
that the density-density and the particle-hole operators are orthogonal to 
the corresponding eigenvectors, and hence the correlation lengths will be 
determined by other eigenvalues.

An interesting question is which local operators $M$ that potentially can 
give true long-range order in the correlation functions Eqs. 
(\ref{eq:expecboson}) and (\ref{eq:expecfermion}). We show in Appendix 
\ref{longrange} that true 
long-range order for bosonic operators is not possible if 
$\hbox{\rm tr} [M^B]= 0$. For fermionic operators we find that true 
long-range order is not possible 
for off-diagonal operators. This explains why there is no true long-range 
order in the density-density and particle-hole correlation functions. The 
proof exploits that the Hamiltonian conserves the number of particles and 
that it is particle-hole symmetric. If we break the particle-hole symmetry, 
we are only guaranteed that the particle-hole correlation function can not 
be truly long-ranged.

Thus, the DMRG will approximate infinite correlation lengths by finite, and 
it is interesting to investigate how the DMRG correlation lengths depend on 
the number of kept states and on the gap of the system.

First of all we need to identify the leading eigenvalues of $\widehat{F}$ 
and $\widehat{1}$ govering the particle-hole and density-density correlations 
respectively.  These eigenvalues can be identified either by a matrix product 
calculation of the respective correlation function, or by measuring the 
correlation length directly in the DMRG calculation. Degeneracies in the 
spectrum are also instrumental in identifying the leading eigenvalue. For 
instance, if we want to compute the particle-hole correlation function, we 
expect this correlation to couple to an eigenvalue that is two-fold 
degenerate since the hole-particle correlation function has an equal 
correlation length. The density-density correlation function, on the other 
hand, will couple to a non-degenerate eigenvalue since there is no symmetry 
related correlation function that demand an equal eigenvalue.

Using the analytical result for the correlation length given by Eq. 
(\ref{eq:phcorrlength}) together with $\lambda =\exp [-1/\xi ]$, we find 
that the exact expression for the eigenvalue dominating the particle-hole 
correlation is:
\begin{equation}
\lambda_{ph}^*(t,\epsilon )= \sqrt{1+\frac{\epsilon^2}{t^2}}-
\frac{\epsilon}{t}.
\label{eq:exacteval}
\end{equation}
The $^*$ is used to indicate that this is an exact value. The expression 
will be used as a reference when we evaluate our numerical data.

\begin{figure}[tbh]
\centerline{\epsfxsize=\columnwidth\epsffile{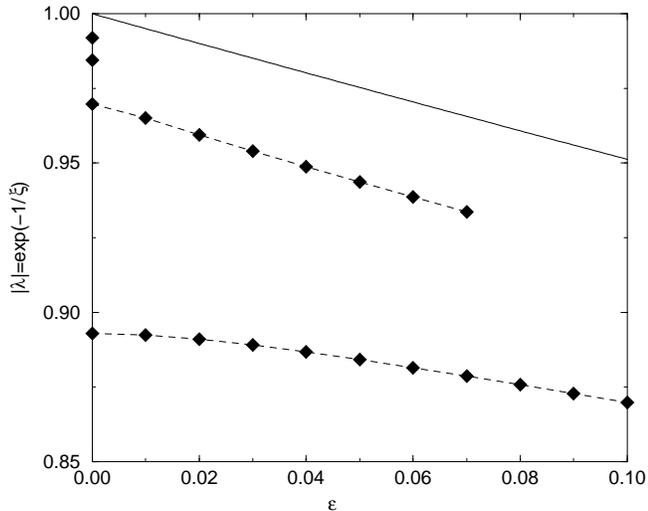}}
\caption[partholeev]{\label{fig:partholeev} The eigenvalues of $\widehat{F}$ 
govering the particle-hole 
correlation length versus $\epsilon$ for different numbers of kept states. 
The solid line corresponds to the exact result from Eq. (\ref{eq:exacteval}) 
From the bottom to the top, the point sets correspond to the following 
numbers of kept states: $(4,6),(12,12),(21,20),(36,32)$. Due to numerical 
problems, we can not continue the set with $m_1$ or $m_2>12$ to larger 
$\epsilon$ values.}
\end{figure}

\begin{figure}[tbh]
\centerline{\epsfxsize=\columnwidth\epsffile{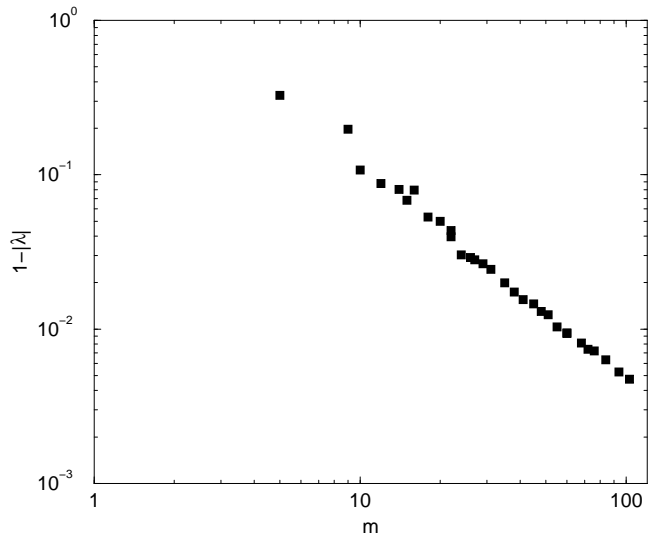}}
\caption[phgapless]{\label{fig:phgapless} Convergence of the eigenvalue 
$\lambda$ of $\widehat{F}$ govering 
the particle-hole correlation length. In the figure we show $1-|\lambda |$ 
versus the number of kept states, $m=m_1+m_2$, in the gapless 
($\epsilon =0$) case.}
\end{figure}

In Fig. \ref{fig:partholeev} our results for the particle-hole correlation 
length are shown. It is clear that as we increase the number of kept states 
in our truncated Hilbert space, the accuracy of the correlation length 
increases. For the case $\epsilon=0$ we see that the eigenvalue 
$|\lambda_{ph}|$ approaches the exact value $1$, i.e. an infinite 
correlation length, as the number of states is increased. The convergence 
of $|\lambda_{ph}|$ towards $1$ is more clearly seen in Fig.
\ref{fig:phgapless}, where we consider the gapless case and plot 
$1-|\lambda_{ph}|$ versus the number of kept states, $m=m_1+m_2$. Thus, 
we conclude that the DMRG gives exponentially decaying correlation functions 
in the gapless case, but as the number of states is increased the correlation 
length grows towards infinity.

We can actually make this conclusion more quantitative. As is seen in Fig.
\ref{fig:phgapless}, the eigenvalue $|\lambda_{ph}|$ behaves as 
$|\lambda_{ph}|\simeq 1-k m^{-\beta}$. Thus the correlation length behaves as 
\begin{equation}
\xi_{ph}\simeq -\frac{1}{\ln |1-k m^{-\beta}|}\simeq \frac{1}{k} m^{\beta}.
\label{eq:scalingformula}
\end{equation}
That is, the correlation length scales as a power of $m$. We find the 
exponent $\beta \simeq 1.3$ and $k\simeq 0.45$. The density-density 
correlation function gives similar results.

In any realistic DMRG computation, the correlation function will be given 
by a sum of a {\em finite} (even if large) number of exponentially decaying 
functions according to Eq. (\ref{eq:corr}). Keeping only a few states, we 
have seen that the correlation function approximates the true power-law 
(Eq. (\ref{eq:partholecorr})) for short correlations, but as we increase $l$ 
we eventually end up with an exponential decay. Increasing the number of 
kept states will make the correlation function look like a power-law for 
rather large $l$, but in the end, when $l\rightarrow\infty$, it will always 
behave as an exponentially decaying function with correlation length given 
by Eq. (\ref{eq:scalingformula}).

%%%%%%%%%%%%%%%%%%%%%%%%%%%%%%%%%%%%%%%%%%%%%%%%%%%%%%%%%%%%%%%%%%%%%%

\subsection{The single-particle excitation spectrum}
\label{bloch}

\begin{figure}[tbh]
\centerline{\epsfxsize=\columnwidth\epsffile{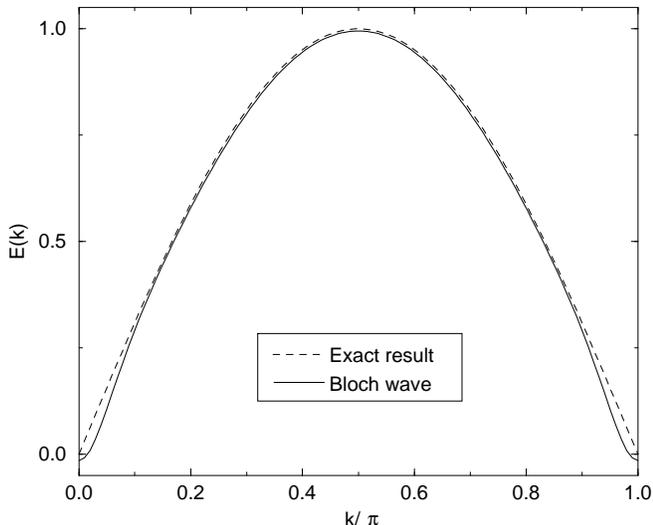}}
\caption[figexcit]{\label{fig:excit} Single particle energy dispersion 
relation $E(k)$. The Bloch-wave 
result deviates from the exact result only close to the Fermi points. 
$(m_1,m_2)=(4,4)$.}
\end{figure}

In order to study the single particle excitation spectrum, we have used the 
Bloch-wave ansatz of Eq. (\ref{eq:bloch}) and the pole-expansion technique
\cite{rommer} to calculate the spectrum. The result is shown in Fig.
\ref{fig:excit}. The curve shows a pair of excitations corresponding to a 
single particle/hole (or spin $S^z=\pm1$). We can see that the dispersion 
relation obtained from the Bloch-wave ansatz is in good agreement with the 
exact dispersion $E(k)=\sin{k}$, except close to the Fermi points, where 
the gap closes. Instead of having a linear form, the calculated dispersion 
relation has the form $E(k)\simeq\Delta_0+v_0^2k^2$ close to $k=0$. 
Furthermore, our excitations have negative energies close to the Fermi 
points, a consequence of a defect ground state. Such a negative energy 
gap was also found\cite{ostlund} to appear for the biquadratic spin-1 chain 
somewhere between the Heisenberg point and the Takhtajan-Babujian point. We 
have investigated how the size of this negative energy gap depends on the 
number of kept states in the truncated Hilbert space. These calculations 
are computationally demanding and also sensitive to numerical errors. 
However, the results we have indicate that the size of the negative energy 
gap decreases as the number of states is increased, but the numerical 
control was poor for $m_1+m_2>12$.

A possible explanation of this defect ground state could be that the DMRG 
has an instability of some sort. We have investigated whether the DMRG at 
$\epsilon =0$ is unstable against breaking the translational symmetry of 
ground state by starting the DMRG calculation with a non-zero staggered 
on-site potential and then, after about 20 iterations, turning the potential 
off. We then let the projection operator converge and compare it with the 
projection operator obtained by a DMRG calculation with the on-site potential 
turned off all the time. However, the two converge to the same limit 
(within the numerical accuracy) with only one exception. If we keep 3 states 
in the integer representation and 6 states in the half-integer 
representation, we find that the DMRG is actually unstable and we obtain 
an energetically more favorable state by breaking the translational 
symmetry. However, it is sufficient to add a single state in the integer 
representation in order to remove this instability. In addition, we have 
also performed variational calculations allowing for a ground state with 
periodicity two, but the energetically lowest state turns out to be 
translationally invariant. Thus, it seems like the DMRG is stable against 
forming a ground state which is not translationally invariant.

There is an interesting symmetry concerning the dispersion relation in Fig.
\ref{fig:excit}. We find that the dispersion relation has the symmetry 
$E(\pi -k)=E(k)$ in the thermodynamic limit. This is in fact a consequence 
of the block structure of $A[s]$, and should therefore be a characteristic 
feature of many half-integer spin systems. To prove this we use the $R$ 
matrix to explicitly construct a Bloch state of momentum $k+\pi$ from a 
state of momentum $k$ and show that these two states have equal energy. 
Thus, we have shown that $E(k)=E(k+\pi )$, but this also proves that 
$E(k)=E(\pi -k)$ since we have the sequence of mappings:
\begin{equation}
E(k)\stackrel{\cal P}{\longrightarrow}E(-k)\stackrel{R}{\longrightarrow}
E(\pi-k),
\label{eq:dispsymm}
\end{equation}
where ${\cal P}$ is a parity transformation, an exact symmetry of our model. 
The details of the proof can be found in Appendix \ref{blochproof}. This 
symmetry is true in general for translationally invariant half-integer spin 
systems with local interactions. As a test, we have checked numerically that 
it is true also for the isotropic antiferromagnetic spin-1/2 Heisenberg 
chain. Furthermore, calculations on the spin-1 chain, as expected, lack 
this symmetry. The symmetry has recently been studied on more general 
grounds by Kladko\cite{kladko}, without any reference to the Bloch-wave 
ansatz. Similarly to our proof for the Bloch-wave ansatz, Kladko explicitly 
constructs a state with momentum $k+\pi$ from a state with momentum $k$ and 
then shows that these states have equal energy.

%%%%%%%%%%%%%%%%%%%%%%%%%%%%%%%%%%%%%%%%%%%%%%%%%%%%%%%%%%%%%%%%%%%%%%
%%%%%%%%%%%%%%%%%%%%%%%%%%%%%%%%%%%%%%%%%%%%%%%%%%%%%%%%%%%%%%%%%%%%%%

\section{Conclusions}\label{conc}

We have investigated fundamental properties of the DMRG when applied to a 
gapless system of free fermions. We find that the DMRG projection operator 
converges to a fixed-point. This convergence means that states of a 
matrix-product form are identical to the DMRG states in the thermodynamic 
limit. By using the matrix-product formalism, we have found that DMRG 
calculations give qualitatively wrong particle-hole and density-density 
correlation functions: the DMRG correlations decay exponentially, while the 
true correlations decay algebraically. However, for short distances, the 
DMRG correlation function agrees with the exact result. The finite 
correlation length of the DMRG particle-hole correlation function scales 
as $\xi \sim m^{1.3}$, where $m$ is the number of kept states. In addition, 
we have derived conditions for whether a general operator potentially can 
give rise to truly long range correlations or not. These conditions are 
found to be determined by symmetries of the Hamiltonian. 

We have demonstrated that the matrix product formalism can be used to 
calculate overlaps between differently renormalized states. This makes 
it possible to directly compare states obtained from DMRG calculations, 
using different numbers of kept states. We propose this overlap as a 
criterion of convergence of DMRG states.

From the matrix-product ansatz we obtain accurate values for the ground 
state energy. Furthermore, using a Bloch-wave ansatz, we find a dispersion 
relation for the excitations that is close to the exact result. Despite 
this, close to the Fermi points, where the gap closes, our excitations 
have negative energies, indicating that the ground state is defect. We 
have not yet been able to trace the origin of these negative energy 
excitations, although our calculations indicate that the magnitude of this 
negative energy gap decreases as the number of kept states is increased. 
In addition, we have shown that the Bloch-wave ansatz for the excitation 
spectrum exhibits the symmetry $E(k)=E(\pi -k)$ for translationally 
invariant half-integer spin systems with local interactions.

%%%%%%%%%%%%%%%%%%%%%%%%%%%%%%%%%%%%%%%%%%%%%%%%%%%%%%%%%%%%%%%%%%%%%%
%%%%%%%%%%%%%%%%%%%%%%%%%%%%%%%%%%%%%%%%%%%%%%%%%%%%%%%%%%%%%%%%%%%%%%

\acknowledgments

The authors would like to thank Stefan Rommer for fruitful discussions. 
We acknowledge the support of the Swedish Research Council for 
Engineering Sciences (TFR) and the Swedish Natural Science Research 
Council (NFR).

%%%%%%%%%%%%%%%%%%%%%%%%%%%%%%%%%%%%%%%%%%%%%%%%%%%%%%%%%%%%%%%%%%%%%%
%%%%%%%%%%%%%%%%%%%%%%%%%%%%%%%%%%%%%%%%%%%%%%%%%%%%%%%%%%%%%%%%%%%%%%

\appendix

\section{Spectrum of $\widehat{F}$}
\label{fhatspec}

Since $\widehat{F}$ determines the possible correlation lengths of fermionic 
operators, it is important to understand the eigenvalue spectrum of this 
operator. We will in this appendix show that the eigenvalues of $\widehat{F}$ 
are related to  those of $\widehat{1}$ by a factor $i$. That is, 
$\lambda_F=i\lambda_1$. To show this, we will start by constructing an 
eigenvector of $\widehat{F}$ with eigenvalue $i$. This construction is 
similar to the one used by Rom\'an {\em et. al.}\cite{roman}. Using 
$F[s,s']=\delta_{s,s'}i^{2s+1}$, we may write the matrix elements of the 
operator $\widehat{F}$ as
\begin{eqnarray*}
&&\widehat{F}^{(\gamma_2',m_2')(\gamma_1',m_1'),(\gamma_2,m_2)
	(\gamma_1,m_1)}\frac{}{}\cr
&&\hspace{0.5cm}=\sum_si^{2(m_2'-m_2)+1}A^{(\gamma_2',m_2')(\gamma_2,m_2)}[s]
A^{(\gamma_1',m_1')(\gamma_1,m_1)}[s]\frac{}{},
\end{eqnarray*}
where we have used that since the projection operator conserves the particle 
number, the element $A^{(\gamma ',m'),(\gamma ,m)}[s]$ is zero unless 
$m'=m+s$. Next, we define the vector 
$|u_F\rangle^{(\gamma_2,m_2)(\gamma_1,m_1)}=\delta_{\gamma_2,\gamma_1}
\delta_{m_2,m_1}i^{2m_2}$. We will now show that $|u_F\rangle$ is an 
eigenvector of $\widehat{F}$ with eigenvalue $i$. We have
\begin{eqnarray*}
&&(\widehat{F}|u_F\rangle )^{(\gamma_2',m_2')(\gamma_1',m_1')}\frac{}{}\cr
   &&\hspace{0.5cm}= \sum_{\gamma_1,\gamma_2}\sum_{m_1,m_2,s}i^{2(m_2'-m_2)+1}
	A^{(\gamma_2',m_2')(\gamma_2,m_2)}[s]\frac{}{}\cr
   &&\hspace{1.0cm}\times A^{(\gamma_1',m_1')(\gamma_1,m_1)}[s]i^{2m_2}
	\delta_{m_1,m_2}\delta_{\gamma_1,\gamma_2}\frac{}{}\cr
   &&\hspace{0.5cm}= ii^{2m_2'}\sum_{\gamma_1,m_1,s}A^{(\gamma_2',m_2')
	(\gamma_1,m_1)}[s]A^{(\gamma_1',m_1')(\gamma_1,m_1)}[s]\frac{}{}\cr
   &&\hspace{0.5cm}= ii^{2m_2'}\delta_{m_2',m_1'}\delta_{\gamma_2',\gamma_1'}
	 = i|u_F\rangle^{(\gamma_2',m_2')(\gamma_1',m_1')}\frac{}{},
\end{eqnarray*}
which proves the claim. In the third line we have used the identity 
$\sum_sA[s]A^T[s]=\openone$. Let us now show that the entire eigenvalue 
spectrum of $\widehat{F}$ is related to that of $\widehat{1}$ by a factor 
$i$. First of all we note that $|\det u_F |=1$, which implies that the 
inverse $u_F^{-1}$ exists. In fact $u_F^{-1}=u_F^{\dagger}$, i.e. $u_F$ 
is unitary. Furthermore, $u_F$ satisfies the equation
\begin{equation}
iu_FA[s]=F[s,s]A[s]u_F,
\label{eq:ufdef}
\end{equation}
since this equation is equivalent to $\widehat{F}|u_F\rangle =
i|u_F\rangle$, as can be seen by simply multiplying Eq. (\ref{eq:ufdef}) 
by $A^{\dagger}[s]$ and summing over $s$. Define the unitary operator 
$\xi =(u_F\otimes 1)$ and consider
\begin{eqnarray*}
i\xi\widehat{1}\xi^{\dagger}
  &=& i(u_F\otimes 1)\biggl( \sum_sA[s]\otimes A[s]\biggr) (u_F^{\dagger}
\otimes 1)\cr
  &=& i\sum_s (u_FA[s]u_F^{\dagger})\otimes A[s]\cr
  &=& \sum_s F[s,s]A[s]\otimes A[s](u_Fu_F^{\dagger}\otimes 1) = \widehat{F}.
\end{eqnarray*}
This implies that $\widehat{F}$ has exactly the same spectrum as 
$i\widehat{1}$, which was our claim. In the third line we have used 
Eq. (\ref{eq:ufdef}).

%%%%%%%%%%%%%%%%%%%%%%%%%%%%%%%%%%%%%%%%%%%%%%%%%%%%%%%%%%%%%%%%%%%%%%
%%%%%%%%%%%%%%%%%%%%%%%%%%%%%%%%%%%%%%%%%%%%%%%%%%%%%%%%%%%%%%%%%%%%%%

\section{Conditions for True Long-Range Order}
\label{longrange}

In this appendix, we show how symmetries of the Hamiltonian determine which 
local operators that potentially can give true long-range order in 
correlation functions. In order for a local bosonic operator $M^B$ to give 
true long-range order it must hold that at least one of the following 
expectation values is non-zero:
$\langle 1|\widehat{M^B}|1\rangle$, $\langle 1|\widehat{M^B}|R\rangle$ and 
$\langle R|\widehat{M^B}|R\rangle$, where $|1\rangle$ and $|R\rangle$ denote 
the eigenvectors of $\widehat{1}$ with eigenvalues $1$ and $-1$ respectively. 
These expectation values will be determined if we can determine the 
expectation values of $A^*[s]\otimes A[s']$ for all combinations of $s$ 
and $s'$. We will frequently interpret the $m^2$ state-vectors $|1\rangle$ 
and $|R\rangle$ as $m\times m$ matrices denoted by $\openone$ and $R$. The 
matrix $R$ has the block-form $\openone\oplus -\openone$ which implies 
that $R^2=\openone$. A subscript $L$ on the matrix denotes that it 
represents the left eigenvector. Furthermore, we will consider the projection 
matrix $A[s]$ to be real.

Let us start by relating different expectation values. We have
\begin{eqnarray*}
&&\langle R|A[s]\otimes A[s']|R\rangle =\hbox{\rm tr} 
	[R_L^TA[s]RA^T[s']]\frac{}{}\cr
&&\hspace{0.5cm}=- \hbox{\rm tr} [R_L^TRA[s]A^T[s']]=-\hbox{\rm tr} 
	[\openone_L^TA[s]A^T[s']]\frac{}{}\cr
&&\hspace{0.5cm}=-\langle 1|A[s]\otimes A[s']|1\rangle \frac{}{},
\end{eqnarray*}
and similarly,
$$
\langle R|A[s]\otimes A[s']|1\rangle =-\langle 1|A[s]\otimes A[s']|R\rangle .
$$

The indices of $A[s]$ correspond to different states and we will label 
these states as $|\gamma ,m\rangle$, where $m$ is the particle number measured 
from half-filling and $\gamma$ is an integer labeling particle-hole 
representations. The transformation of the state $|\gamma ,m\rangle$ under a 
particle-hole transformation ${\cal B}$ is
\begin{equation}
{\cal B}|\gamma ,m\rangle =\phi(\gamma ,m)|\gamma ,-m\rangle ,
\end{equation}
where $\phi (\gamma ,m)=\pm 1$. That is, the state $|\gamma ,m\rangle$ 
transforms within the $\gamma$ representation of the particle-hole symmetry 
group. Note also that ${\cal B}^2=1$, which implies that $\phi$ is 
independent of $m$, since ${\cal B}^2|\gamma ,m\rangle =\phi (\gamma ,m)
\phi (\gamma ,-m)|\gamma ,m\rangle$.

We are now going to show that $\openone_L^{(\gamma ',m'),(\gamma ,m)}$ is 
zero unless $m'=m$. To see this we write the equation $\langle 1|=\langle 1|
\widehat{1}$ as
\begin{eqnarray*}
\openone_{L}^{(\gamma_1 ,m_1),(\gamma_2,m_2)}&=&\sum_s\sum_{\gamma_3,
\gamma_4,m_3,m_4} \openone_{L}^{(\gamma_3 ,m_3),(\gamma_4,m_4)}\frac{}{}\cr
	&&\times A^{(\gamma_3,m_3),(\gamma_1,m_1)}[s]A^{(\gamma_4 ,m_4),
	(\gamma_2,m_2)}[s]\frac{}{}.
\end{eqnarray*}
Using the fact that $A$ conserves the particle number (see Appendix 
\ref{fhatspec}), we conclude that $m_1-m_2=m_3-m_4$. Since the difference 
between the $m$-values of $\openone_L^{(\gamma_1 ,m_1),(\gamma_2,m_2)}$ is 
conserved under the action of $\widehat{1}$, we can write $\openone_L=
\openone_L^0+\openone_L^{rest}$ where $\openone_L^0$ only has non-zero 
matrix elements between states of equal particle number and 
$\openone_L^{rest}$ is the remainder, mixing particle numbers. This 
decomposition will not be mixed under the action of $\widehat{1}$ and 
hence both matrices must be eigenstates of $\widehat{1}$ with eigenvalue 
$1$. This eigenvalue is, however, non-degenerate, which means that either 
of $\openone_L^0$ and $\openone_L^{rest}$ is zero. Since 
$\hbox{\rm tr} \openone_L=\langle 1|1\rangle =1$, $\openone_L$ must contain 
$\openone_L^0$ and hence $\openone_L^{rest}=0$. This means that 
$\openone_L$ only connects states containing an equal number of particles.

Furthermore, since $R_L=\openone_LR$ and $R$ also conserves the number of 
particles (it is diagonal), we conclude that $R_L$ conserves the number of 
particles. 

Let us now consider the following expectation value:
\begin{eqnarray*}
&&\langle 1|A[s]\otimes A[s']|1\rangle=\hbox{\rm tr}
	[\openone_L^TA[s]A^T[s']]\frac{}{}\cr
&&\hspace{0.5cm}=\sum_{\{\gamma_i,m_i\}}(\openone_L^T)^{(\gamma_1 ,m_1),(
	\gamma_2, m_2)}\frac{}{}\cr
&&\hspace{1.0cm}\times A^{(\gamma_2,m_2),(\gamma_3,m_3)}[s]A^{(\gamma_1,m_1),(
	\gamma_3,m_3)}[s']\frac{}{}.
\end{eqnarray*}
Using particle number conservation of $A$ and $\openone_L$ we conclude 
that the expectation value is zero unless $s=s'$, i.e.
\begin{equation}
\langle 1|A[s]\otimes A[s']|1\rangle \sim \delta_{s,s'}.
\label{eq:deltaeqa}
\end{equation}
Similarly, using $R_L$ instead of $\openone_L$, one finds
\begin{equation}
\langle 1|A[s]\otimes A[s']|R\rangle\sim\delta_{s,s'}.
\label{eq:deltaeqb}
\end{equation}
To summarize, Eqs. (\ref{eq:deltaeqa}) and (\ref{eq:deltaeqb}) follow from the 
particle-number conserving property of the Hamiltonian together with the 
uniqueness of the eigenvalue $1$ of $\widehat{1}$. In the next paragraph, 
we will investigate the expectation values when $s=s'$. Note that the Pauli 
matrix $\sigma_3$ and $\openone$ form a complete basis for all diagonal 
operators $M^B$, which according to the above results are the only operators 
that can give true long-range order.

Recall the defining relation of the projection operator:
\begin{equation}
|\gamma ',m'\rangle =\sum_{(\gamma ,m),s_j}A^{(\gamma ',m'),(\gamma ,m)}[s_j]
|\gamma ,m\rangle\otimes |s_j\rangle .
\label{eq:defrel}
\end{equation}
Applying the particle-hole transformation to the defining relation 
Eq. (\ref{eq:defrel}), we find that $A$ has the following symmetry:
\begin{equation}
A^{(\gamma ',m'),(\gamma ,m)}[s_j]=\phi (\gamma ')\phi (\gamma )(-1)^j
A^{(\gamma ',-m'),(\gamma ,-m)}[-s_j].
\label{eq:aphsym}
\end{equation}
Let us write down the operator form of $A[s]$
$$
A[s_j]=\sum_{(\gamma ',m'),(\gamma ,m)}A^{(\gamma ',m'),(\gamma ,m)}[s_j]
|\gamma ',m'\rangle\langle\gamma ,m |.
$$
Applying a particle-hole transformation to $A[s]$, making use of the 
symmetry Eq. (\ref{eq:aphsym}), one finds that
\begin{equation}
{\cal B}A[s_j]{\cal B}=(-1)^jA[-s_j].
\end{equation}
Using the definition of $\widehat{1}$ we find the transformation property 
of $\widehat{1}$ under particle-hole transformations:
$$
({\cal B}\otimes {\cal B})\sum_sA[s]\otimes A[s]({\cal B}\otimes {\cal B})
	=\sum_sA[-s]\otimes A[-s].
$$
That is, $\widehat{1}$ is invariant under particle-hole transformations. Let 
us also consider the transformation of $\widehat{\sigma_3}$, where $\sigma_3$ 
is a Pauli matrix. Using exactly the same technique, we find that 
$({\cal B}\otimes {\cal B})\widehat{\sigma_3}({\cal B}\otimes {\cal B})=
-\widehat{\sigma_3}$.

$\langle 1|$ and $\langle 1|({\cal B}\otimes {\cal B})$ must both have 
eigenvalue 
$1$ of $\widehat{1}$, since $\widehat{1}$ is invariant under particle-hole 
transformations. Using the non-degeneracy of this eigenvalue, we conclude 
that $\langle 1|({\cal B}\otimes {\cal B})=e^{-i\theta} 
\langle 1|$ and similarly 
$({\cal B}\otimes {\cal B})|1\rangle =e^{i\theta}|1\rangle$.

Let us now compute the expectation value of $\widehat{\sigma_3}$,
$$
\langle 1|\widehat{\sigma_3}|1\rangle =
\langle 1|({\cal B}\otimes {\cal B})^2
\widehat{\sigma_3}({\cal B}\otimes {\cal B})^2|1\rangle =-
\langle 1|\widehat{\sigma_3}|1\rangle ,
$$
and thus $\langle 1|\widehat{\sigma_3}|1\rangle =0$. Since $\langle 1|
\widehat{1}|1\rangle =1$ we arrive at
\begin{equation}
\langle 1|A[s]\otimes A[s']|1\rangle =-\langle R|A[s]\otimes A[s']|R\rangle
=\frac{1}{2}\delta_{s,s'}.
\end{equation}

It only remains to consider expectation values of $\widehat{1}$ and 
$\widehat{\sigma_3}$ between $\langle 1|$ and $|R\rangle$. Trivially, 
$\langle 1|\widehat{1}|R\rangle =0$, since $\langle 1|$ and $|R\rangle$ are 
eigenvectors of $\widehat{1}$ with different eigenvalues. Using the 
structure of $|1\rangle$ and $|R\rangle$, let us show that they must transform 
with the same phase factor. Noting that ${\cal B}$ must be block-diagonal, 
it follows that $[{\cal B},R]=0$. Assuming $({\cal B}\otimes {\cal B})
|1\rangle={\cal B}{\cal B}^T=e^{i\theta}|1\rangle$ we find
$$
({\cal B}\otimes{\cal B})|R\rangle ={\cal B}R{\cal B}^T=R{\cal B}{\cal B}^T=
Re^{i\theta}\openone =e^{i\theta}|R\rangle .
$$
Using this property we find
$$
\langle 1|\widehat{\sigma_3}|R\rangle =-
\langle 1|\widehat{\sigma_3}|R\rangle =0
$$
and we conclude
\begin{equation}
\langle 1|A[s]\otimes A[s']|R\rangle =-
\langle R|A[s]\otimes A[s']|1\rangle =0.
\end{equation}

Using the derived expressions for the expectation values, a general bosonic 
operator $\widehat{M^B}$ can couple to the eigenstates with corresponding 
eigenvalues $\pm 1$ only if $\hbox{\rm tr} M^B\neq 0$. In particular, the 
density-density correlation function can not be truly long-ranged when the 
Hamiltonian has particle-hole symmetry and conserves the number of particles.

Let us also investigate correlation functions between fermionic operators, 
$M^F$, in order to be able to draw conclusions concerning the particle-hole 
correlation function. We will only show that off-diagonal fermionic 
operators can not give rise to truly long-ranged correlation functions. 
From Eq. (\ref{eq:expecfermion}) we see that true long-range order is possible 
only if some expectation value of the form 
$\langle 1|\widehat{S^+}|u_F \rangle$ is non-zero, that is $\widehat{S^+}$ 
must connect eigenstates of $\widehat{1}$ and $\widehat{F}$ with the 
corresponding eigenvalues having absolute value $1$. We will now show that 
this is impossible due to the particle-number conserving property of $A[s]$. 
Note that all the important eigenvectors are proportional to
$\delta_{m,m'}$. Hence, an expectation value of the above form will be
\begin{eqnarray*}
\langle 1|\widehat{S^+}|u_F\rangle &\sim& \sum\delta_{m,m'}\delta_{m,1/2+m''}
\delta_{m',-1/2+m'''}\delta_{m'',m'''}\frac{}{}\cr
&=&0\frac{}{},
\end{eqnarray*}
where the second and third delta-functions come from the particle-number 
conserving property of $A[s]$. Thus we can not have true long-range order 
in the particle-hole correlation function.  

Note that we have not proved that the correlation function between two 
traceless fermionic operators (like $\sigma_3$) can not be long ranged. 
On the contrary, this is the structure of the string order correlation 
function in the spin-1 chain, which is long-ranged.

%%%%%%%%%%%%%%%%%%%%%%%%%%%%%%%%%%%%%%%%%%%%%%%%%%%%%%%%%%%%%%%%%%%%%%
%%%%%%%%%%%%%%%%%%%%%%%%%%%%%%%%%%%%%%%%%%%%%%%%%%%%%%%%%%%%%%%%%%%%%%

\section{Proof of the Bloch-state Symmetry}
\label{blochproof}

In this appendix, we prove that the energy spectrum obtained from the 
Bloch-state ansatz exhibits the symmetry $E(k)=E(\pi -k)$. Since the 
symmetry $E(k)=E(-k)$ follows from parity being a good quantum number, 
we only need to show that $E(k)=E(k+\pi )$. The strategy used in the proof 
is to construct a Bloch-state of momentum $k+\pi$ from a state of momentum 
$k$ and to show that expectation values of these two states are equal in 
the thermodynamic limit.

Let us first recall some properties of the eigenvector $R$ of $\widehat{1}$. 
We will use the convention that $|R\rangle$ is an $m^2$ vector and $R$ is an 
$m\times m$ matrix and similarly we will write the eigenvector of 
$\widehat{1}$ corresponding to the eigenvalue $1$ as $|1\rangle$ or 
$\openone$. Using the block-diagonal structure of $R$ it follows that 
$\{ (\openone\otimes R),\widehat{M}\} =0$ and that 
$[(R\otimes R),\widehat{M}]=0$ for all local operators $M$.

We also need the property $(R\otimes R)\lim_{n\rightarrow\infty}
\widehat{1}^n=\lim_{n\rightarrow\infty}\widehat{1}^n$. To show this we 
recall that all but two eigenvalues of $\widehat{1}$ have absolute value 
less than one. The corresponding eigenvectors will be annihilated by 
$\lim_{n\rightarrow\infty}\widehat{1}^n$. Hence we may write
$$
\lim_{n\rightarrow\infty}\widehat{1}^n=|1\rangle\langle 1|
+(-1)^n|R\rangle\langle R|.
$$
If we let this operator act on a general state 
$|\psi\rangle =\sum_i\psi_i|i\rangle$ we obtain
$$
\lim_{n\rightarrow\infty}\widehat{1}^n|\psi\rangle =\psi_1|1\rangle +(-1)^n
\psi_R|R\rangle .
$$
Now, if we act on the resulting vector with $(R\otimes R)$, we obtain 
(now we use the matrix form of the vectors)
\begin{eqnarray*}
(R\otimes R)\lim_{n\rightarrow\infty}\widehat{1}^n|\psi \rangle
&=&(R\otimes R)[\psi_1|1\rangle +(-1)^n\psi_R|R\rangle ]\frac{}{}\cr
&=&\psi_1R\openone R^T+(-1)^n\psi_RRRR^T\frac{}{}\cr
&=&\psi_1\openone+(-1)^n\psi_RR =\lim_{n\rightarrow\infty}
	\widehat{1}^n|\psi\rangle \frac{}{}.
\end{eqnarray*}
Thus we have shown that $(R\otimes R)\lim_{n\rightarrow\infty}\widehat{1}^n$ 
and $\lim_{n\rightarrow\infty}\widehat{1}^n$ act equally on a general state 
and hence the operators must be identical.

We are now ready to show that for all local operators $M$, it holds that
\begin{eqnarray}
\hbox{LHS}&=&(RQ',k|M|RQ,k)\frac{}{}\cr
	&=&(Q',k+\pi |M|Q,k+\pi )=\hbox{RHS}\frac{}{},
\label{eq:exequal}
\end{eqnarray}
where the Bloch-states are defined in Eq. (\ref{eq:bloch}). If we apply 
this result to the Hamiltonian operator and the normalization, we obtain 
the result that the states $|RQ,k)$ and $|Q,k+\pi )$ have equal energy. 
That is, we have a mapping from a state of momentum $k$ to a state of 
momentum $k+\pi$ with equal energy, which proves the symmetry 
$E(k)=E(k+\pi )$. 

In order to prove Eq. (\ref{eq:exequal}), we begin by writing the Bloch-state 
$|RQ,k)$ as
\begin{eqnarray*}
|RQ,k)&=&\sum_{j,\{ s_j\}}e^{ij(k+\pi)}\hbox{\rm tr} \bigl[ RA[s_N]
\cdots A[s_{j+1}]Q\cdots A[s_1]\bigr]\frac{}{}\cr
	&&\times |s_N\cdots s_1\rangle\frac{}{} ,
\end{eqnarray*}
where we have used $\{ R,A[s]\} =0$ to move the $R$ to the left side of the 
trace. Thus we may write\cite{rommer} the left-hand side of Eq. 
(\ref{eq:exequal}) as
\begin{eqnarray*}
\hbox{LHS}&=&\sum_{j,j'}e^{ij(k+\pi)}e^{-ij'(k+\pi)}\times\hbox{\rm tr} 
\bigl[ (R\otimes R)\widehat{1}^{N-j}\frac{}{}\cr
	&&\times (\openone\otimes Q)\widehat{1}^{j-l}
	\widehat{M}\widehat{1}^{l-j'-1}(Q'^*\otimes\openone )
	\widehat{1}^{j'}\bigr]\frac{}{} ,
\end{eqnarray*}
where we have assumed that $M$ acts on the site $l$. Now, when we go to 
the thermodynamic limit $N\rightarrow\infty$, we will always have a factor 
$\widehat{1}^{\infty}$ somewhere in the trace. If we could move 
$(R\otimes R)$ through the trace until it reaches the $\widehat{1}^{\infty}$ 
factor, the $(R\otimes R)$ would be annihilated and we would be left with 
an expression that is
\begin{eqnarray*} 
&&\sum_{j,j'}e^{ij(k+\pi)}e^{-ij'(k+\pi)}\hbox{\rm tr} 
	\bigl[\widehat{1}^{N-j}(\openone\otimes Q)\frac{}{}\cr
&&\hspace{0.5cm}\times
	\widehat{1}^{j-l}\widehat{M}\widehat{1}^{l-j'-1}(Q'^*
\otimes\openone )\widehat{1}^{j'}\bigr] =\hbox{RHS}\frac{}{}
\end{eqnarray*}
and our proof would be complete. It turns out that it is always possible 
to perform such a move. If the factor $\widehat{1}^{\infty}$ is not between 
the factors $(\openone\otimes Q)$ and $(Q'^*\otimes\openone)$, we just 
commute $(R\otimes R)$ through the $\widehat{\hspace{5pt}}$-operators 
until it reaches $\widehat{1}^{\infty}$ and gets annihilated. If the 
factor $\widehat{1}^{\infty}$ is between $(\openone\otimes Q)$ and 
$(Q'^*\otimes\openone)$ we can make the split $(R\otimes R)=
(\openone\otimes R)(R\otimes\openone )$ and move these factors in 
different directions until they meet between $(\openone\otimes Q)$ and 
$(Q'^*\otimes\openone)$ and get annihilated by $\widehat{1}^{\infty}$. 
During the movements we will pick up a factor $(-1)^{j''+(N-j'')}$, and 
since we are only considering chains of even length, this factor is equal 
to $1$. Note that the proof works even if the operator $M$ does not act on 
a single site, but, as for example the Hamiltonian, acts on a couple of 
neighboring sites. The important thing is that the operator is local, so 
that a factor of $\widehat{1}^{\infty}$ always can be found in the trace. 

This completes the proof.

%%%%%%%%%%%%%%%%%%%%%%%%%%%%%%%%%%%%%%%%%%%%%%%%%%%%%%%%%%%%%%%%%%%%%%
%%%%%%%%%%%%%%%%%%%%%%%%%%%%%%%%%%%%%%%%%%%%%%%%%%%%%%%%%%%%%%%%%%%%%%

\end{multicols}


\begin{references}
\bibitem{hamsign}
We have chosen to include a factor $-1$ in the hopping term of the 
Hamiltonian in order to have a translationally invariant ground state, 
but this is only a matter of convenience. By simply making the canonical 
transformation $c_j\rightarrow (-1)^jc_j$, we can remove this sign from $H$.
\bibitem{energydensity}
In the thermodynamic limit $N\rightarrow\infty$, the ground state energy 
per site is given by 
$$
E_0=-\frac{|\epsilon|}{\pi} E\Bigl[ -\frac{t^2}{\epsilon^2},
\frac{\pi}{2}\Bigr] ,
$$
where $E$ is an elliptic integral of the second kind.
\bibitem{psi anal} The result is derived by introducing periodic
boundary conditions, letting $N\rightarrow\infty$, and studying the
asymptotic behavior of $C(l)$ for large $l$.
\bibitem{evalpair}
From the block-structure of $A[s]$ it follows that $\widehat{M}$ has 
the block-form
$$
\widehat{M}=\pmatrix{ 0 & M_a \cr
		      M_b & 0}.
$$
Assuming that $u=u_a\oplus u_b$ is an eigenvector of $\widehat{M}$ with 
corresponding eigenvalue $\lambda$, it follows that $M_au_b=\lambda u_a$ 
and $M_bu_a=\lambda u_b$. Defining $\bar{u}=u_a\oplus -u_b$ we find
$$
\widehat{M}\bar{u}=\pmatrix{M_au_b \cr -M_bu_a}=\pmatrix{\lambda u_a \cr 
-\lambda u_b}=-\lambda \bar{u},
$$
i.e. $\bar{u}$ is an eigenvector of $\widehat{M}$ corresponding to the 
eigenvalue $-\lambda$.
\bibitem{nonsymmetry}
Generally the matrix $\widehat{1}$ is non-symmetric, which forces us to 
distinguish between right and left eigenvectors.
\end{references}
\end{document}